\documentclass[conference]{IEEEtran}
\IEEEoverridecommandlockouts
\usepackage{cite}
\usepackage{amsmath,amssymb,amsfonts}
\usepackage{algorithmic}
\usepackage{graphicx}
\usepackage{textcomp}
\usepackage{xcolor}
\def\BibTeX{{\rm B\kern-.05em{\sc i\kern-.025em b}\kern-.08em
    T\kern-.1667em\lower.7ex\hbox{E}\kern-.125emX}}
\begin{document}

\title{xIDS-EnsembleGuard: An Explainable Ensemble Learning-based Intrusion Detection System\\ {\footnotesize Presented at TrustCom-2024, the 23rd IEEE International Conference on Trust, Security and Privacy in Computing and Communications, China.} \thanks{Acknowledgement: This work was supported in part by the Guangdong Provincial Key Laboratory of Human Digital Twin (Grant 2022B1212010004), Guangzhou Basic Research Program (Grant SL2023A04J00930), and the Shenzhen Holdfound Foundation Endowed Professorship
} }


\author{\IEEEauthorblockN{1\textsuperscript{st} Muhammad Adil}
\IEEEauthorblockA{\textit{Deptartment of Computer Science } \\
\textit{University at Buffalo}\\
Buffalo, USA \\
muhammad.adil@ieee.org }
\and
\IEEEauthorblockN{2\textsuperscript{nd} Mian Ahmad Jan}
\IEEEauthorblockA{\textit{College of Computing and Informatics} \\
\textit{University of Sharjah}\\
Sharjah, UAE \\
mjan@sharjah.ac.ae}
\and
\IEEEauthorblockN{3\textsuperscript{rd} Safayat Bin Hakim}
\IEEEauthorblockA{\textit{School of Information Systems} \\
\textit{University of Maryland}\\
Baltimore County, USA \\
shakim3@umbc.edu}

\and 
\IEEEauthorblockN{4\textsuperscript{th} Houbing Herbert Song}
\IEEEauthorblockA{\textit{School of Information Systems} \\
\textit{University of Maryland}\\
Baltimore County, USA \\
songh@umbc.edu}
\and

\IEEEauthorblockN{5\textsuperscript{th} Zhanpeng Jin}
\IEEEauthorblockA{\textit{School of Future Technology} \\
\textit{South China University of Technology }\\
Guangdong, China, And \\
\textit{University at Buffalo }\\
Buffalo, USA \\
zjin@scut.edu.cn}}


\maketitle

\begin{abstract}
In this paper, we focus on addressing the challenges of detecting malicious attacks in networks by designing an advanced Explainable Intrusion Detection System (xIDS). The existing machine learning and deep learning approaches have invisible limitations, such as potential biases in predictions, a lack of interpretability, and the risk of overfitting to training data. These issues can create doubt about their usefulness, and transparency, and decrease the trust of involved stakeholders. To overcome these challenges, we propose an ensemble learning technique called the "EnsembleGuard". This approach uses the predicted outputs of multiple models, including tree-based (LightGBM, GBM, Bagging, XGBoost, CatBoost) and deep learning models such as neural network (LSTM (long short-term memory networks) and GRU (gated recurrent unit), to maintain a balance and achieve trustworthy results.  Our work is unique because it combines both tree-based and deep learning models to design an interpretable and explainable meta-model through model distillation. By considering the predictions of all individual models, our neta-model effectively addresses key challenges, and ensures both explainable and reliable results.
We evaluate our model using well-known datasets, including UNSW-NB15, NSL-KDD, and CIC-IDS-2017, to assess its reliability against various types of attacks.  During analysis, we found that our model outperforms both tree-based models and other comparative approaches when it comes to different kinds of attack scenarios.

\end{abstract}


\section{Introduction}
In today's fast-changing digital world, Next Generation Networks (NGN) are revolutionizing connectivity and data exchange among interconnected devices. Moreover, these networks have become an integral part of our daily lives. Therefore, it is important to ensure the security of these networks against cyber threats \cite{b1}. The nature of new threats is changing from general to more sophisticated, which highlights the need for more advanced Intrusion Detection Systems (IDSs) that could be capable of identifying and mitigating emerging security challenges \cite{b2}.  However, the deployment of such an IDS could be challenging due to the dependency on black-box  models, which lack transparency and interpretability \cite{b3}. The complex nature of these models makes it hard to see how they make decisions and to identify or rectify any biases in their predictions \cite{b4}. Similarly, this lack of clarity limits security professionals' understanding regarding certain detected attacks, which is important for improving security measures \cite{b5}. Another big problem with these models is the lack of compatibility with new attacks \cite{b6}. Furthermore, there is a constant risk of overfitting the training data. When this occurs, the system may become too familiar with the training data and incorrectly classify legitimate traffic as security vulnerabilities that are not actually present. This could make the IDS less reliable when it is used in real-world situations \cite{b7, b8}.

In response to these challenges, the academic and industry communities have explored various techniques to improve the reliability of IDS. However, these approaches have certain limitations that undermine the trust of the stakeholders involved. Therefore, there is an urgent need to develop an advanced and intelligent IDS that can overcome these limitations without losing the generality. In this paper, we present the EnsembleGuard Meta model, a unique approach that combines the outputs of both tree-based and deep-learning models. This model aims to tackle challenges such as potential biases in predictions, lack of interpretability, poor generalization to unseen data, and the risk of overfitting to the training data. Moreover, it provides a balanced, scalable, and reliable solution for IDS by effectively handling new or unseen data. We will give a brief overview of the model's architecture and operational steps to set the stage for a detailed exploration.

Moreover, the key contributions of this work are summarized as below.

\begin{enumerate}
    \item Initially, we implement and evaluate tree-based models like LightGBM, GBM, Bagging, XGBoost, and CatBoost, along with deep learning models such as LSTM and GRU. We used the UNSW-NB15 \cite{b9}, NSL-KDD \cite{b10}, and CIC-IDS2017 datasets \cite{b11} for their evaluation..
    \item EnsembleGuard is different from state-of-the-art ensemble learning techniques because it uses the prediction output of different tree-based and deep learning-based models in the meta-model, which helps to overcome issues such as potential biases in predictions, lack of interpretability, poor generalization to unseen data, and the risk of overfitting to the training data followed by the explainability of detected attacks.
    \item Next, we experimentally demonstrate that EnsembleGuard outperforms similar models constructed from tree-based models and other comparative models when it comes to different data evaluations.
    \item Finally, we evaluate EnsembleGuard using various comparison metrics to see its results in the presence of comparative approaches. We also look into how well this model detects different types of attacks to show its unique contribution.

\end{enumerate}

\textbf{Paper Structure:} In Section II, we covered the related work associated with IDS. In Section III, we talk about the motivation for this work and the limitations of the state-of-the-art schemes, while Section IV, covers the proposed model and the experimental setup. Section V presents the results and statistics, and Section VI evaluates the EnsembleGuard Meta-model against rival algorithms to highlight its superior performance in terms of consistency and accuracy across different datasets. Finally, Section VII summarizes and concludes the paper.

\section{Related Work}

In recent years, the deployment of IDS for protecting NGN has become a key focus in cybersecurity research. With the help of machine learning and deep learning algorithms, these systems can identify unusual traffic patterns in the network that may indicate cyber threats. While these models can effectively detect illegal traffic, however, sometimes it can be challenging to understand the underline decision-making process. This complexity makes it difficult for security experts to understand, interpret, and trust how the security threats have been detected and identified. Because of this, it is important to point out the key areas where the IDS decision-making process can be improved. 

Zhou et al. \cite{b12} proposed an intelligent IDS using the feature selection processes of different classifiers, while Saleh et al. \cite{b13} introduced a hybrid IDS that integrates a prioritized k-nearest Neighbors (kNN) algorithm with optimized Support Vector Machines (SVM) classifiers. In \cite{b14}, the authors focused on web-based attacks by proposing an advanced IDS for NGNs. In \cite{b15}, the author designed a dataset to analyze various targeted attacks, while Karimi et al. \cite{b16} proposed a KNN-based IDS to address the security issues in networks.

Zhang et al. \cite{b17} introduced the IDS model known as the "IDSSTG2P" framework that effectively addresses data imbalance through the borderline synthetic minority oversampling technique. In addition, they implemented K-means clustering to enhance classification accuracy and reduce the False Acceptance Rate (FAR). Similarly, Yao et al. \cite{b18} developed an intelligent IDS model that also tackles data imbalance issues. They streamlined the feature set by applying an autoencoder (AE) to reduce the dimensions of the data to classify the refined features using the LightGBM algorithm.

In \cite{b19}, the author proposed an advanced IDS using Deep Belief Networks (DBNs) alongside Restricted Boltzmann Machines (RBMs) to resolve the security problems of networks, while, Yang et al. \cite{b20} explored the effectiveness of DBNs for intrusion detection using the KDDCup99 dataset.

In \cite{b21}, the authors proposed a DL-enabled IDS for resource-limited networks. Moreover, they claimed that their proposed model achieves higher accuracy rates compared to traditional machine learning-enabled IDS.  Maddu et al. \cite{b22} also introduced a DL-enabled IDS for Software-Defined Networking (SDN) to address the security challenges that arise with single-point failure due to targeted attacks. However, this model is evaluated for limited attacks, which keeps its practicability in a fuzzy state. In \cite{b23}, Saikam et al. proposed an IDS framework that combined the power of DL and sampling techniques to address security problems in resource-limited networks. The goal of this approach was to address the challenges posed by imbalanced data and improve the accuracy of attack detection. Moreover, we have added Table \ref{tab:contributions} in the paper to summarize the contributions of different techniques that have been used in the recent past to design robust IDS.

\begin{table*}[htbp]

\centering

\caption{Existing State-of-the-art techniques and their results statistics}
\label{tab:contributions}
\begin{tabular}{|p{2.3cm}|p{1.1cm}|p{4.9cm}|p{2.1cm}|p{2.2cm}|p{2cm}|}
\hline
 \textbf{\tiny Reference} & \textbf{\tiny Year of Publication } & \textbf{\tiny Contribution} & \textbf{\tiny Dataset Used} & \textbf{\tiny Result Accuracy} & \textbf{\tiny Uniqueness} \\
\hline

Elnakib et al. \cite{b24} & 2023 & Deep Learning Multiclass classification model (EIDM) to detect different attacks in live networks.  &  CICIDS2017  & 95\% & Checked on one dataset\\

\hline

Qazi et al. \cite{b25} & 2023 & The authors used a Hybrid Deep-Learning-Based Network Intrusion Detection System (HDLNIDS) by employing a convolutional recurrent neural network. & CICIDS2017 & 98\% & Checked on one dataset\\
\hline 
Awajan et al. \cite{b26} &  2023 & DL-enabled IDS for Internet of Things applications. & Not discussed & 93.74\% & Checked on one dataset \\
\hline 
Chaganti et al. \cite{b27} & 2023 & LSTM-based IDS for IoT applications  & SDNIoT-focused datasets &  97.1\%. & Checked on one dataset \\
\hline 
Balla et al. \cite{b28} & 2024 & The authors used a hybrid deep learning model of CNN-LSTM to design an intelligent IDS.  &  CICIDS2017 & 95.5 \% & Checked on one dataset \\
\hline

Musthafa  et al. \cite{b29} & 2024 & To mitigate overfitting issues, they designed LSTM-based IDS for operational networks.   & UNSW-NB15 & 97.23\% & Checked on one dataset \\
\hline

Ashiku et al. \cite{b30} & 2021 & DL-enabled IDS for multiple attack detection & UNSW-NB15 & 97-99\% & Checked on one dataset\\

\hline 

 EnsembleGuard & 2024 & For the first time, we used tree-based and Deep Learning models to design EnsembleGuard Meta-model & UNSW-NB15, NSL-KDD and CIC-IDS-2017 & All datasets $\approx$  98.8\% & Checked on three datasets\\

\hline
\end{tabular}
\end{table*}

\section{Motivation of this work and Limitations of State-of-the-Art-Schemes}

The development of robust xIDS has become an utmost requirement of networks. Although there have been notable strides in improving these systems because many existing IDS) still face challenges. These challenges not only limit their practical use across different applications but also impact their reliability and effectiveness in detecting complex threats. Our proposed EnsembleGuard Meta model aims to address these challenges by leveraging the collective strengths of different machine learning models, which facilitates explainable and trustworthy results. The motivation behind this was the observation of the existing IDS, where most of them are evaluated on a limited dataset, which can lead to potential biases, lack of interpretability, poor generalization to unseen data, and the risk of overfitting, etc. By combining multiple models through an ensemble approach, our Meta-model mitigates these limitations and offers the undermentioned key advantages in presence of state-of-the-art models.

\begin{itemize}
    \item \textit{Improve Accuracy and Robustness:} The EnsembleGuard, which combines multiple models that can capture various aspects of the data and help offset each model's weaknesses. This collaborative approach of tree-based models such as LightGBM, GBM, Bagging, XGBoost, CatBoost, and neural network models such as LSTM, GRU ultimately enhances the overall attack detection accuracy and robustness of the meta-model.
    
    \item \textit{Interpretability and Explainability:} Our Meta-model purifies the output of the multi-models by combining them into a single interpretable model that can provide understandable results and enable a better sense of the underlying patterns of attacks.\\
    The interpretability part is very important because it helps the network administrators and security analysts to check the attack ratio and type.

    \item \textit{Generalization and Adaptability:} The Meta-model demonstrates its ability to detect a wide range of attack types and adapt to different network environments, as it is trained and tested on different datasets. \\
    The generalization capability of the meta-model ensures its effectiveness for real-world deployment, where network traffic patterns and attack vectors are constantly changing.
    \item \textit{Standardized Framework:} The Meta-model serves as a standardized framework for network intrusion detection by providing consistent and interpretable results that can be easily adopted into the existing security paradigm of the networks.  \\
    
\end{itemize}

Furthermore, our evaluation results show that the EnsembleGuard Meta model performs better than both the individual algorithms used in this study and other comparison models. This highlights its effectiveness and strength in detecting various types of attacks. To summarize, our work addresses the limitations of existing IDS solutions by improving the accuracy, interpretability, and generalization in the network.

\section{Proposed Model}

In this section, we introduce the EnsembleGuard Meta model, a novel approach designed to detect malicious traffic and attacks within networks. This model harnesses the collective strengths of various tree-based and deep-learning techniques with the objective of improving detection accuracy and consistency. The core idea behind the EnsembleGuard Meta model is to combine the predictive capabilities of traditional tree-based models with those of deep learning (DL) models, which improves the overall performance and robustness of IDS. This framework uses a range of well-established algorithms, such as LightGBM, GBM, XGBoost, and CatBoost, as well as deep learning algorithms like LSTM and GRU. Initially, we train these models independently and subsequently aggregate their predictions within the meta-model. This aggregation allows the model to capture a more comprehensive representation of the underlying data patterns and attack signatures present in the network.

This meta-model acts as a decision-making layer by combining the knowledge distilled from the considered tree-based model and DL models. The final predictions made by the Meta-model leverage the collective intelligence of the underlying models, which leads them to improved accuracy, robustness, and generalization capabilities of IDS in the network. The unique aspect of the EnsembleGuard Meta model lies in its ability to provide explainable and trustworthy results in terms of specific attack detection metrics. While testing a decision tree-based meta-model, the system can offer interpretable insights into the decision-making process, which allows network administrators and security analysts to understand the rationale behind the intrusion detection results. This interpretability is important for building confidence in the system's performance and facilitating its integration into real-world security infrastructures.

\begin{figure*}[hbt]
	\centering
		\includegraphics[width=.75\linewidth, height = 6.5 cm ]{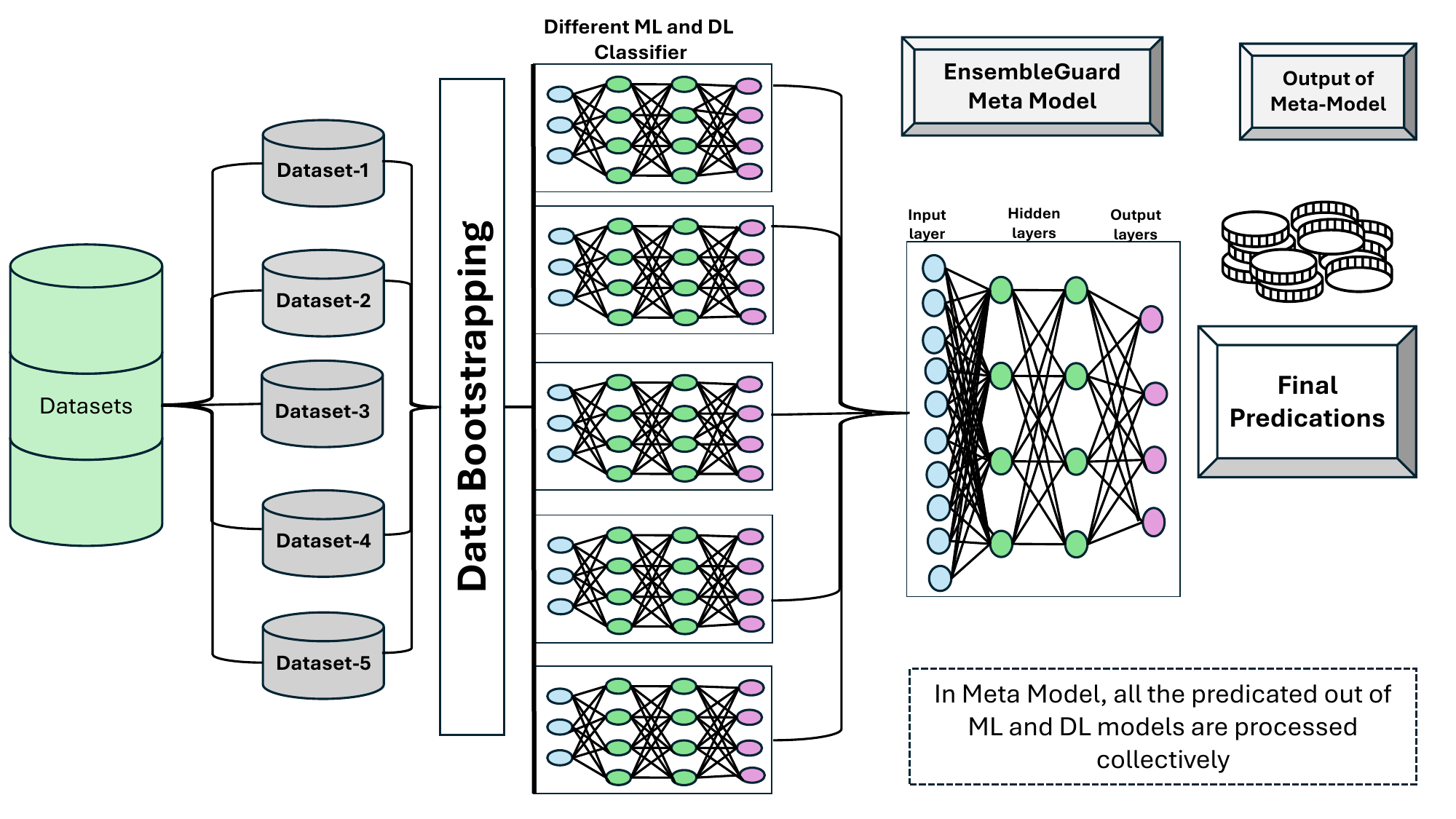}
		\caption{Visual Illustration of the Proposed  EnsembleGuard Meta-Model Structure }
		\label{fig : 01} 
	
\end{figure*}

To provide a hypothetical overview of the proposed meta-model, we included Figure \ref{fig : 01} within the paper. This figure illustrates a comprehensive view of our methodology. On the left side of the figure, there is a large dataset box, which is subdivided into five distinct subdatasets, denoted as dataset-1 through dataset-5. These sub-datasets represent the hypothetical overview of the dataset considered in this work for training and testing both tree-based and deep-learning models. Following the data bootstrapping step, these models generate outputs, which are then used as inputs for the Ensemble Guard meta-model. The Ensemble Guard meta-model comprises its own input, hidden, and output layers, functioning to predict the final output. This approach ensures a holistic understanding of our methodology's intricacies and facilitates the comprehension of our research framework.

\subsection{Experiment Setup}

In this section, we discuss the data management, testing, and evaluation of the proposed model. First, we carefully worked with different datasets like UNSW-NB15, NSL-KDD, and CIC-IDS-2017, to prepare them for the training and testing. Next, we tested tree-based models, deep learning models, and meta-models both individually and together to understand and highlight the unique contributions of the meta-model. We compared the performance of these models using various metrics to determine which ones performed best, including our proposed meta model.

\subsection{Data Cleaning}

In this section, we discuss how different datasets have been cleaned. Initially, all the datasets are thoroughly checked for missing values using the imputation technique. To generalize this process, let's denote the original datasets as $D = \{(x_i, y_i)\}_{i=1}^{n}$, where $x_i \in \mathbb{R}^{p}$ represents the feature vector and $y_i \in \mathbb{R}$ represents the target variable for the $i^{\text{th}}$ instances in $D$.

Next, we identified the instances with missing values in the dataset and simplified it with the notation $D_{\text{m}} = \{(x_i, y_i) \in  D \quad | \quad \exists, \quad j, x_{ij} = \text{Na}\}$. Consequently, we applied the imputation technique to handle missing values, $\hat{x}_{ij} = f_{\text{imp}}(x_i, D \setminus D_{\text{m}})$. Thereafter, we update the dataset such that $D = D \setminus D_{\text{m}} \cup \{(\hat{x}_i, y_i)\}_{i=1}^{|D_{\text{m}}|}$. Next, we determined the potential outliers using a statistical technique, which is generalized as $D_{\text{out}} = \{(x_i, y_i) \in D | |x_i - \mu| > k\sigma\}$, where $\mu$ and $\sigma$ are the mean and standard deviation of the feature vector, and $k$ is a hyperparameter. 
 Following that, we identify the categorical features, which are simplified by $\mathcal{C} = \{j | x_{ij} \in \text{categorical}\}$, and apply label encoding to the categorical features such that $\hat{x}_{ij} = f_{\text{label-encode}}(x_{ij})$ for $j \in \mathcal{C}$. Thereafter, we update $D = \{(\hat{x}_i, y_i)\}_{i=1}^{n}$ to complete the data cleaning process. Now, the dataset $D$ is transformed into a high-quality, well-structured input for the meta-model training and testing. Next, we randomly split the dataset into training and testing sets such that $D_{\text{tr}} = {(x_i, y_i)}{i=1}^{n{\text{tr}}}$, $D_{\text{te}} = {(x_i, y_i)}{i=n{\text{tr}}+1}^{n}$, respectively, where $n_{\text{tr}}$ is the number of instances in the training set, and $n - n_{\text{tr}}$ is the number of instances in the testing set.

 \subsection{GBM and LightGBM classifiers Implementation}

In this section, we discuss various tree-based machine-learning techniques that have been used in the implementation of our meta-model. Initially, we explored the simple Gradient Boosting Machine (GBM) and LightGBM classifiers. We trained and tested these classifiers separately to extract their results. Following this, we utilized the predicted outputs from these models as inputs for our meta-model, which we designed as EnsembleGuard. This approach allowed us to leverage the strengths of each classifier and enhance our overall model performance.

\subsection{Bagging, XGB, and CatBoot Classifier Implementation}

In this section, we discuss three more tree-based classifiers, which are named, the Bagging classifier, XGBoost (XGB), and CatBoost. For the Bagging classifier, we begin with initialization and set the base estimator to a DecisionTreeClassifier, with the number of estimators (decision trees) configured to 1000. Next, we apply the simple training and testing approach for both the XGB and CatBoost classifiers. This process allows us to extract the predicted outputs, which are then used as inputs in our meta-model, alongside the predicted outputs from the Bagging classifier to design EnsembleGuard.

\subsection{LSTM and GRU Models Implementation}

In this section, we discuss the Long Short-Term Memory (LSTM) and Gated Recurrent Unit (GRU) deep learning models by focusing on how they have been implemented and utilized in the proposed meta-model. Let's first examine the LSTM model, which consists of three layers: an input layer with 128 units, a dropout layer with a rate of 0.2 to introduce regularization and help mitigate overfitting, and an output layer with a number of unique classes corresponding to the target variable Y. The activation function used for the output layer is set to 'softmax' to facilitate multi-class classification. The output of the LSTM then serves as input for the meta-model. To train the neural networks, we utilize the 'adam' optimizer, while the cross-entropy loss function is selected to effectively handle multi-class classification tasks. We followed the similar structure for the GRU model to maintain consistency across the input, hidden, and output layers. Just as with the LSTM, the predicted outputs of the GRU are also fed into the meta-model. This systematic approach ensures that both models contribute effectively to the overall performance of the meta-model.

\section{Performance Evaluations of Tree based Model and Deep Learning Models}

In this section, we provide a comprehensive analysis of various tree-based models, including GBM, LightGBM, Bagging, CatBoost, and XGBoost, as well as deep learning models like LSTM and GRU. These models are checked on the UNSW-NB15, NSL-KDD, and CIC-IDS-2017 datasets. Firstly, we evaluate the performance of these models individually using key metrics such as precision, recall, F1-score, and accuracy. These metrics help us assess how effectively each model can detect different types of attacks in the datasets. This thorough evaluation give us valuable insights into the strengths and weaknesses of each approach in addressing network intrusion detection challenges.

\subsection{Performance Evaluation based on UNSW-NB15}

We initially assessed the performance of all tree-based models on the UNSW-NB15 dataset and achieved results ranging from 95\% to 98\% for different metrics. These results demonstrate the effectiveness of the models in classifying each attack in the test dataset. However, our primary intention was to apply these outputs to our proposed model to enhance the validity and trustworthiness of our findings.
In the next phase, we analyzed the performance of deep learning models and observed impressive results varying from 96\% to 98.5\%. This comparison underscores the fact that different model paradigms yield different results, which support our argument of designing a comprehensive model that maintains the trust of all stakeholders. Given that, we designed EnsembleGuard, which was also checked, and we achieved results between 98\% and 99\%.  The results obtained during these experiments are shown in Table \ref{tab:bagging_classifier}.

\begin{table*}[htbp!]
\centering
\caption{Result Statistics of Tree-based Classifier, Deep Learning Model and Meta-Model for UNSW-NB15 Dataset }
\begin{tabular}{l | c | c | c | c}
\hline
\multicolumn{5}{c}{\textbf{ GBM, LightGBM, Bagging, XGB and CatBoost Classifier Classifier Results stastistics}} \\
\hline
\textbf{Different Classes}& \textbf{Precision Results} & \textbf{Recall Score} & \textbf{F1-score Results} & \textbf{Support} \\

\hline
\multicolumn{5}{c}{\textbf{ GBM Classifier Result Statistics}} \\

\hline

\textbf{ Accuracy Results} & 0.981 & 0.978 & 0.975 & 175341 \\

\hline
\multicolumn{5}{c}{\textbf{ Bagging Classifier Result Statistics}} \\

\textbf{ Accuracy Results} & 0.978 & 0.979 & 0.981 & 175341 \\

\hline
\multicolumn{5}{c}{\textbf{ XGB  Classifier Result Statistics}} \\
\hline
\textbf{ Accuracy Results} & 0.958 & 0.971 & 0.965 & 175341 \\

\hline
\multicolumn{5}{c}{\textbf{  CatBoost Classifier Result Statistics}} \\
\hline

\textbf{ Accuracy Results} & 0.979 & 0.983 & 0.978 & 175341 \\

\hline

\multicolumn{5}{c}{\textbf{  GRU Model Result Statistics Result Statistics}} \\
\hline

\textbf{ Accuracy Results} & 0.983 & 0.975 & 0.986 & 175341 \\

\hline 

\multicolumn{5}{c}{\textbf{  LSTM Model Result Statistics Result Statistics}} \\
\hline

\textbf{  Accuracy Results} & 0.971 & 0.988 & 0.979 & 175341 \\

\hline

\multicolumn{5}{c}{\textbf{  EnsembleGuard Meta-model Result Statistics Result Statistics}} \\
\hline

\textbf{ Accuracy Results} & \textbf{0.986} & \textbf{0.988} & \textbf{0.986} & 175341 \\
\hline

\end{tabular}
\label{tab:bagging_classifier}
\end{table*}

\subsection{Performance Evaluation based on CIC-IDS-2017 Dataset}

In this section, we discuss the statistical results obtained from evaluating both tree-based models and deep learning models on the dataset of CIC-IDS-2017. For this analysis, these models were tested on a dataset encompassing multiple classes, each representing different types of attacks such as DoS attacks, web attacks, botnet attacks, port scans, brute force attacks, and infiltration, etc. The diversity of these attack classes highlights the complexity of the dataset and the challenges it presents to design an IDS that can effectively detect each of them with results statistics in the network. Therefore, we realized and ensure to in our meta-model paradigm to show the attacks detection results statistics with explainability. To ensure a fair comparison, we employed the same evaluation metrics for all models. In addition, we incorporated an individual attack scenario into our analysis. This approach enabled us to closely examine how effectively each model responds to both seen and unseen attacks, thereby providing a more cleared understanding of their performance.  The results of this evaluation are presented in Table \ref{tab:dl-based-222}.

\begin{table}[htbp!]
\tiny
\centering
\caption{Comparative  Result Statistics of EnsembleGuard and Its sub-models for Multi-class classification on CIC-IDS-2017 Dataset  }
\begin{tabular}{l | c | c | c | c}
\hline
\hline
\textbf{Different Classes}& \textbf{Precision Results} & \textbf{Recall Score} & \textbf{F1-score Results} & \textbf{Support} \\
\hline
\multicolumn{5}{c}{\textbf{  LightGBM Classifier Result Statistics for }} \\
\hline
         DoS  Attacks &     0.94  &    0.92    &  0.92    &   387 \\
   WebAttack   Attacks &   0.94    &  0.91    &  0.93    &    14 \\
         Botnet  Attacks &   0.85    &  0.88   &   0.81   &    612\\
    PortScan   Attacks  & 0.82   &   0.842    &  0.79   &      8 \\
  BruteForce    Attacks &  0.93    &  0.89   &   0.92     &  231\\
Infiltration   Attacks &   0.88   &   0.81   &   0.85   &    452 \\

\hline
\multicolumn{5}{c}{\textbf{   Bagging Classifier Result Statistics for }} \\
\hline
         DoS  Attacks &     0.93  &    0.94    &  0.93    &   387 \\
   WebAttack   Attacks &   0.91    &  0.94    &  0.89    &    14 \\
         Botnet  Attacks &   0.92    &  0.95   &   0.92   &    612\\
    PortScan   Attacks  & 0.80   &   0.81    &  0.77   &      8 \\
  BruteForce    Attacks &  0.82    &  0.80   &   0.86     &  231\\
Infiltration   Attacks &   0.86   &   0.84   &   0.83   &    452 \\

\hline

\multicolumn{5}{c}{\textbf{ XGB Classifier  Result Statistics for  }} \\
\hline
       DoS  Attacks &     0.90  &    0.92    &  0.87    &   387 \\
   WebAttack   Attacks &   0.86    &  0.86    &  0.91    &    14 \\
         Botnet  Attacks &   0.84    &  0.89   &   0.87   &    612\\
    PortScan   Attacks  & 0.75   &   0.77    &  0.73   &      8 \\
  BruteForce    Attacks &  0.82    &  0.88   &   0.85     &  231\\
Infiltration   Attacks &   0.83   &   0.90   &   0.87   &    452 \\
\hline

\multicolumn{5}{c}{\textbf{ CatBoost  Model Result Statistics  }} \\

\hline

       DoS  Attacks &     0.89 &    0.90    &  0.94    &   387 \\
   WebAttack   Attacks &   0.90    &  0.93    &  0.91    &    14 \\
         Botnet  Attacks &   0.94    &  0.95   &   0.97   &    612\\
    PortScan   Attacks  & 0.77   &   0.75    &  0.80   &      8 \\
  BruteForce    Attacks &  0.93    &  0.95   &   0.92     &  231\\
Infiltration   Attacks &   0.92   &   0.90   &   0.94   &    452 \\
\hline

\multicolumn{5}{c}{\textbf{ LSTM  Model Result Statistics }} \\

\hline

       DoS  Attacks &     0.95 &    0.95    &  0.96    &   387 \\
   WebAttack   Attacks &   0.96    &  0.92    &  0.95    &    14 \\
         Botnet  Attacks &   0.91    &  0.96   &   0.96   &    612\\
    PortScan   Attacks  & 0.90   &   0.92    &  0.85   &      8 \\
  BruteForce    Attacks &  0.96    &  0.98   &   0.97     &  231\\
Infiltration   Attacks &   0.98   &   0.97   &   0.98   &    452 \\
\hline

\multicolumn{5}{c}{\textbf{ GRU  Model Result Statistics  }} \\

\hline

       DoS  Attacks &     0.94 &    0.98    &  0.95    &   387 \\
   WebAttack   Attacks &   0.97    &  0.95    &  0.97    &    14 \\
         Botnet  Attacks &   0.92       &  0.91   &   0.94   &    612\\
    PortScan   Attacks  & 0.92   &   0.88    &  0.90   &      8 \\
  BruteForce    Attacks &  0.97    &  0.96   &   0.96     &  231\\
Infiltration   Attacks &   0.96   &   0.98   &   0.97   &    452 \\

\hline

\multicolumn{5}{c}{\textbf{ EnsembleGuard Meta-model  Result Statistics  }} \\

\hline

       DoS  Attacks &     \textbf{0.98} &    \textbf{0.986}    &  \textbf{0.979}    &   387 \\
   WebAttack   Attacks &   \textbf{0.989}    &  \textbf{0.993}    &  \textbf{0.987}    &    14 \\
         Botnet  Attacks &   \textbf{0.986}       &  \textbf{0.979}   &   \textbf{0.973}   &    612\\
    PortScan   Attacks  & \textbf{0.982}   &   \textbf{0.98}    &  \textbf{0.96}   &      8 \\
  BruteForce    Attacks &  \textbf{0.991}    &  \textbf{0.987}   &   \textbf{0.983}     &  231\\
Infiltration   Attacks &   \textbf{0.986}   &   \textbf{0.986}   &   \textbf{0.982}   &    452 \\

\hline

\end{tabular}
\label{tab:dl-based-222}
\end{table}

\subsection{Performance Evaluation based on NSL-KDD Dataset}

In this section, we present the results of different models evaluated on the NSL-KDD dataset. The primary objective is to assess the performance of these models and understand how they respond to different datasets, thereby examining the correlation of the results. A key concern in this research is that existing models are often not evaluated on multiple datasets simultaneously, which make them less practical for real deployment. In addition, it remains unclear how these models perform when trained on one dataset and tested on other dataset. To address this issue, we evaluated all models on the NSL-KDD dataset, and the results are summarized in Table \ref{tab:dl-based-233}.
Through this comprehensive evaluation, we aim to get insights regarding the behavior and generalization capabilities of these models. Moreover, we learn and understand how these models respond to unseen scenarios and identify any potential limitations or biases. In addition, we evaluate their performance when integrated into our meta model to determine their consistency in the context of results. This analysis provide clarity on the overall performance consistency of our EnsembleGuard framework. Table \ref{tab:dl-based-233} below presents the detailed results of this experimental evaluation, and enables us to compare the performance of the various models and draw meaningful conclusions about their suitability for real-world applications.

\begin{table}[htbp!]
\tiny
\centering
\caption{Result Statistics for NSL-KDD Dataset for Multi-class classification}
\label{tab:dl-based-233}
\begin{tabular}{|l | c | c | c |}
\hline
\textbf{Different Classes} & \textbf{Precision Results} & \textbf{Recall Score} & \textbf{F1-score Results} \\
\hline
\multicolumn{4}{c}{\textbf{LightGBM Classifier Result Statistics}} \\
\hline
DoS Attacks & 0.93 & 0.90 & 0.935 \\
Probing Attacks & 0.84 & 0.88 & 0.85 \\
Privilege Attacks & 0.923 & 0.943 & 0.921 \\
Access Control Attacks & 0.856 & 0.88 & 0.875 \\
\hline
\multicolumn{4}{c}{\textbf{Bagging Classifier Result Statistics}} \\
\hline
DoS Attacks & 0.925 & 0.931 & 0.882 \\
Probing Attacks & 0.962 & 0.953 & 0.967 \\
Privilege Attacks & 0.94 & 0.952 & 0.946 \\
Access Control Attacks & 0.952 & 0.942 & 0.953 \\
\hline
\multicolumn{4}{c}{\textbf{XGB Classifier Result Statistics for Unseen Data}} \\
\hline
DoS Attacks & 0.95 & 0.924 & 0.931 \\
Probing Attacks & 0.91 & 0.94 & 0.89 \\
Privilege Attacks & 0.92 & 0.95 & 0.92 \\
Access Control Attacks & 0.921 & 0.899 & 0.902 \\
\hline
\multicolumn{4}{c}{\textbf{CatBoost Model Result Statistics for Unseen Data}} \\
\hline
DoS Attacks & 0.925 & 0.904 & 0.915 \\
Probing Attacks & 0.926 & 0.917 & 0.931 \\
Privilege Attacks & 0.914 & 0.947 & 0.903 \\
Access Control Attacks & 0.919 & 0.926 & 0.912 \\
\hline
\multicolumn{4}{c}{\textbf{LSTM Model Result Statistics}} \\
\hline
DoS Attacks & 0.86 & 0.81 & 0.84 \\
Probing Attacks & 0.91 & 0.87 & 0.90 \\
Privilege Attacks & 0.88 & 0.86 & 0.87 \\
Access Control Attacks & 0.885 & 0.851 & 0.893 \\
\hline
\multicolumn{4}{c}{\textbf{GRU Model Result Statistics}} \\
\hline
DoS Attacks & 0.92 & 0.952 & 0.942 \\
Probing Attacks & 0.963 & 0.953 & 0.951 \\
Privilege Attacks & 0.939 & 0.936 & 0.953 \\
Access Control Attacks & 0.952 & 0.943 & 0.951 \\
\hline
\multicolumn{4}{c}{\textbf{EnsembleGuard Meta-model Results (for Seen and Unseen Data)}} \\
\hline
DoS Attacks &  0.971 &  0.981 &  0.979 \\
Probing Attacks &  0.974 &  0.971 &  0.976 \\
Privilege Attacks & 0.968 &  0.977 &  0.977 \\
Access Control Attacks &  0.976 &  0.973 &  0.973 \\

\hline
\end{tabular}
\end{table}

\section{Performance Evaluation of EnsembleGuard with Rival Schemes}

In this section, we discuss the statistical results of the EnsembleGuard Meta-model in presence of rival schemes to demonstrate how consistence it is, when it comes to the evaluation of different dataset. Despite the fact, we have previously presented the results of various tree-based models and deep learning approaches, we observed inconsistencies in their evaluations. Now it time to see the results statistics of the state-of-the-art schemes in comparison of our Meta-model.

\begin{table*}[htbp]
\centering
\caption{Why EnsembleGuard, Comparative  Analysis with Rival Schemes }
\begin{tabular}{|p{3cm} | p{4cm} | p{4cm} | p{3cm} |} 
\hline
\textbf{Model/Scheme} & \textbf{Evaluated on Dataset} & \textbf{Limitations} & \textbf{Results} \\
\hline
Elnakib et al. \cite{b31} & CICIDS2017 &  inconsistent results observation with our implemented algorithms  &   only classification based results \\
\hline
Qazi et al. \cite{b32} & CICIDS2017 & inconsistent results observation with our implemented algorithms &   only classification based results \\
\hline
Awajan et al. \cite{b33} & Private Dataset & iinconsistent results observation with our implemented algorithms &  only classification based results \\
\hline
Chaganti et al. \cite{b34} & SDN IoT-focused datasets &  inconsistent results observation with our implemented algorithms &  only classification based results \\
\hline
Balla et al. \cite{b35} & CICIDS2017 &  inconsistent results observation with our implemented algorithms &  only classification based results \\
\hline
Musthafa et al. \cite{b36} & UNSW-NB15 &  inconsistent results observation with our implemented algorithms &  only classification based results \\
\hline
Ashiku et al. \cite{b37} & UNSW-NB15 &  inconsistent results observation with our implemented algorithms &  only classification based results \\
\hline
\textbf{EnsembleGuard } & 
CICIDS-2017, NSL-KDD \& UNSW-NB15 & \textbf{consistent results } &  \textbf{attack based results} \\
\hline
\end{tabular}
\label{tab:dl-based-244}
\end{table*}

In the relevant literature, we noticed inconsistencies in the reported results, as different models achieves differing levels of accuracy on the same dataset. For example, Elnakib et al. \cite{b31} reported an accuracy of 95\% on the CICIDS2017 dataset, while Qazi et al. \cite{b32} achieved a higher accuracy of 98\% using a different model on the same dataset. Similarly, Awajan et al. \cite{b33} used a deep learning-enabled model and reported an accuracy of 93.74\% on a private dataset. Chaganti et al. \cite{b34} developed a LSTM-based IDS, and reported 97.1\% results, while Balla et al. \cite{b35} used CNN-LSTM-based IDS to address the security problem in next-generation network (NGN) networks. They showed 95.5\% accucy for their model. Moreover, Musthafa et al. \cite{b36} designed an enhanced LSTM-based IDS, and reported  97.23\% results on UNSW-NB15 dataset. Similarly, Ashiku et al. \cite{b37} extend this idea by designing a LSTM based IDS for network traffic analysis and reported 98\% results.

Furthermore, our own evaluation of the tree-based and deep learning models revealed similar inconsistencies in the results. These varying results may raise doubts in mind of consumers regarding the reliability of each model and its suitability for real-world deployment. Therefore, we designed a collaborative model in the form of EnsembleGuard that address these issues. Furthermore, to support this point, we have compiled the obtained results of all models in Tables 2-4.
In contrast to the results of individual models, the EnsembleGuard meta-model maintains consistent performance across different datasets, as shown in Tables 2-4. The performance of our meta-model is further compared against rival schemes in Table \ref{tab:dl-based-244}, to highlight why it is necessary in the presence of state-of-the-art IDS.
The consistent results of our meta-model addresses the concerns noted in the literature regarding the lack of consistency in IDS. It provides a more reliable and trustworthy solution for real-world network security applications. In addition, our model displays the percentage of each detected attack, which enhances network security by helping organizations to determine the security measures needed to counter specific threats. This transparency also empowers security professionals to better understand how effectively various attacks have been identified within operational networks.

\section{Conclusion}

In this work, we propose a meta-model known as "EnsembleGuard," which combines the strengths of tree-based models and deep learning models. In the literature review and our experiment results reveal that the existing IDS models often exhibit varying levels of accuracy and reliability, which undermine user trust and confidence in their deployment. 
To address this challenge, the EnsembleGuard meta-model uses the model distillation approach to integrate the output of these models by maintaining a balance between their predicted output that yields more consistent and trustworthy results. The comprehensive evaluations of our meta-model on the UNSW-NB15, NSL-KDD, and CIC-IDS-2017 datasets demonstrate its superiority over rival schemes in terms of accuracy and consistency. Furthermore, the EnsembleGuard meta-model not only achieves high detection rates but also enables users to understand the specific attack ratios. This feature is important for organizations to implement appropriate preventive measures to secure their networks. By tackling issues of consistency and explainability, the EnsembleGuard meta-model paves the way for it broader adoption in many applications.



\vspace{12pt}
\color{red}

\end{document}